\documentclass[fleqn,twoside,twocolumn,nofootinbib]{revtex4} % Specifies the document class %,unsortedaddress
\usepackage{ujp} % \usepackage[cyr]{ujp} for cyrillic, \usepackage[web]{ujp} for web
\begin{document}
\title[STABILITY OF TRIONIC STATES IN ZIGZAG CARBON NANOTUBES]%колонтитул
{STABILITY OF TRIONIC STATES IN ZIGZAG CARBON NANOTUBES}%
\author{S.~Marchenko}%1 автор
\affiliation{Department of Theoretical Physics, I.I.~Mechnikov Odesa National University}%институт
\address{2,~Dvorianska Str., Odesa 65026, Ukraine}%адрес
\email{Sergey.Marchenko@onu.edu.ua}%e-mail
\affiliation{Department of Biophysics, Informatics, and Medical
Equipment,\\ Odesa National
Medical University}%
\address{2, Valikhovskyi Lane, Odesa 65026, Ukraine}%
\udk{} \pacs{61.48.De; 71.35.Pq;\\[-3pt] 73.21.-b} \razd{\secvii}

\setcounter{page}{1055}%
\maketitle

\begin{abstract}
The stability of trionic excitations in zigzag carbon nanotubes has been
estimated. A trion is shown to be unstable with respect to the
ground excitonic state and stable with respect to the excited one.
So, trions in nanotubes of this type can be formed by capturing an
electron or a hole by an excited exciton. In other words, the trion
in a nanotube is an excimer complex, which results in the formation of
a system with three energy levels (unexcited exciton--trion--excited
exciton).
\end{abstract}

\section{Introduction}

Owing to their unique mechanical, optical, and electric properties,
single-walled carbon nanotubes (SWNTs) have attracted much attention
in various domains of researches within the last two decades \cite{jorio1}.
One of the most interesting features of carbon nanotubes is a strong
correlation between charge carriers, which manifests itself through the
quantum confinement; this phenomenon is observed in one-dimensional (1D)
structures about 1~nm in diameter.

The electron-electron repulsion and the electron-hole attraction, which
play an important role in governing the electronic and optical properties of
nanotubes \cite{ando1}, give rise to the formation of excitons with huge
binding energies in semiconducting carbon nanotubes \cite{wang1} and even in
metallic carbon nanotubes \cite{wang2}.

\section{Excitonic States}

The simplest excitation in a semiconducting system consists in that an
electron, after having absorbed energy, transits from the valence band into
the conduction one. In this case, there emerges a hole in the valence band,
which behaves as a positively charged particle. For instance, an electron
and a hole, while interacting by means of only the Coulomb mechanism, can form
bound states. Such states are called excitons. In general, two types of
excitons are distinguished: Frenkel and Wannier--Mott excitons.

Carbon nanotubes belong to the class of direct-band-gap semiconductors.
Therefore, the annihilation of electron-hole pairs can take place in them,
which is accompanied by a photon emission \cite{perebeinos1}. Excitons
localized in carbon nanotubes have an additional restriction; namely, the
nanotube diameter is fixed and, therefore, such excitons can be considered
as inherently one-dimensional objects.

The Coulomb interaction is known to be considerably strengthened in quasi-1D
systems. This circumstance enhances the stability of exciton-like
excitations, which makes the existing Coulomb blockade much stronger.
Together with strong polarization effects, all this results in that new
excitons are hardly formed. In other words, a few excitons, being formed in
a nanotube, could block the formation of new ones, so that the total exciton
concentration may be scarce. The issue concerning the influence of the screening
by excitons on the formation of new many-particle excitations was described
in work \cite{adamyan1}.

The formation of excitons in carbon nanotubes is possible, if the energy
that generates the electron-hole pair is equal or smaller than the energy
gap width. In the case where the energy is sufficient for the electron to
get into the conduction band and for the hole, respectively, into the valence
one, both quasiparticles do not interact, remaining independent of each
other.

In 1D nanotubes, the exciton radius is much larger than the
lattice constant, so that those excitons are analogs of
Wannier--Mott excitons taking place in three-dimensional crystals.
The exciton mass equals the reduced mass of the electron and the
hole,
%1
\begin{equation}
\mu =\frac{m_{e}m_{h}}{m_{e}+m_{h}},  \label{1.1}
\end{equation}%
where $m_{e}$ and $m_{h}$ are the effective masses of the electron and the hole,
respectively. The specific value of effective mass depends on the nanotube
crystal structure. It can be calculated consistently using expressions from
work \cite{tishchenko1}. Below, the interaction between the electron and the
hole will be considered as purely Coulombic.

However, there arises a question: What should be taken as the dielectric
permittivity? For a carbon nanotube, the effects of the screening by its charges
manifest themselves only at distances of the order of the tube diameter.
Therefore, as the first approximation, we may adopt that that $\varepsilon = 1$
for nanotubes. We note that if the nanotube is located in a certain medium,
e.g., in micelles, the corresponding value of dielectric permittivity for
this medium should be accepted, which would affect the energies of excitonic
excitations.

In this work, in order to estimate the energies of exciton-like
excitations, we used the well-known Ritz method [7]. Recall that
this method can be applied to evaluate the characteristic energy
values from above. Note that we are interested in the stability of
excitonic excitations with respect to one-electron states and the
stability of trionic states with respect to the decay into an
exciton and a hole (an electron). Therefore, if calculations are
carried out in the framework of the same model, even this crude
technique will produce qualitatively valid results.

To estimate the characteristic energy values, we use the Schr\"{o}dinger
equation written down in the form
%2
\begin{equation}
\left[ -\frac{\hbar ^{2}}{2\mu }\frac{\partial ^{2}}{\partial
x^{2}}-\frac{e^{2}}{\sqrt{x^{2}+d^{2}}}\right] \Psi =E\Psi.
\label{1.2}
\end{equation}%
Here, $\mu $ is the reduced exciton mass, $e$ the elementary charge, $x$ the
distance between the electron and the hole measured along the nanotube, and
$d$ is a parameter, which depends on the nanotube diameter and, in the first
approximation, is equal to it. One should pay attention that the potential
in this equation is an even function. Therefore, the excitonic states in
nanotubes are divided into two series, each composed of either even or odd
characteristic wave functions. The exciton spectrum for the odd-function
series is similar to those, which are typical of other hydrogenic systems.
Optical transitions are allowed between the states in the even- and
odd-function series. Taking into account that, in the odd-state series, the
wave functions vanish at $x=0$, the characteristic energy values for those
states can be found as characteristic values of Eq.~(\ref{1.2}) determined
at the semiaxis $(0,+\infty )$ with the boundary condition
%3
\begin{equation}
\Psi (0)=0.  \label{1.3}
\end{equation}%
Analogously, the spectrum for the even-function series is determined by the
boundary condition
%4
\begin{equation}
\Psi ^{\prime }(0)=0.  \label{1.4}
\end{equation}

The Hamiltonian in Eq.~(\ref{1.2}) can be used, when the dielectric
permittivity of the medium, in which the nanotube is located, is close to
1. The application of a cut-off Coulomb potential in the model is
associated with the necessity to remove the unphysical singularity at the zero
point, which stems from the one-dimensionality of the problem. However, if one
changes to a consequent consideration of a nanotube as a cylinder, this
potential can be used as the basis for a more exact potential expressed via
an elliptic integral, which was done, e.g., in work \cite{smyrnov1}.

The main criterion for the choice of trial functions in the method used is
their correspondence to boundary conditions. We also demand that they have
to be integrable and smooth. However, it is not a rigorous requirement, but
would only simplify the calculations. The behavior of the selected exciton
wave functions completely coincides with that of exact functions presented
graphically and described in
\cite{dresselhaus1}.

We used the trial functions of the following forms:
%5
\begin{equation}
\Psi _{\rm even}(x)=C_{1}\exp {[-\alpha x^{2}]}  \label{1.5}
\end{equation}for the even-function series and
%6
\begin{equation}
\Psi _{\rm odd}(x)=C_{2}x\exp {[-\alpha x^{2}]}  \label{1.6}
\end{equation}%
for the odd-function one. It is necessary to get rid of dimensional
coefficients. Therefore, let us change to the system of atomic units, in
which $\mu =1$, $\hbar =1$, and $e=1$. The Bohr radius of an exciton serves as
a unit length. Evidently, it depends on the effective mass of an exciton and
the nanotube diameter. Therefore, the nanotube diameter in the final
equation must be expressed in terms of introduced atomic units.

In this case, the ground state in the even-function series approximately
coincides with the minimum of the function
%7
\begin{equation}
I_{1}(\alpha)=\frac{\hbar^{2}\sqrt{2\pi\alpha}}{8\mu}-\frac{1}{2}%
e^{2}e^{\alpha d^{2}}{K}_{0}[\alpha d^{2}],   \label{1.7}
\end{equation}
and, in the odd-function series, with the minimum of the function
%8
\begin{equation}
I_2(\alpha)=\frac{\hbar^{2}\sqrt{2\pi\alpha}}{14\alpha\mu}+\frac{1}{8}\frac{%
e^{2}\sqrt{\pi}}{\alpha}{U}\Big[\frac{1}{2},0,2\alpha d^{2}\Big].
\label{1.8}
\end{equation}
Here, $K_{n}(x)$ is the modified second-kind Bessel function, and
$U(a,b,z)$ is the confluent hypergeometric function.

%Fig. 1
\begin{figure*}% figure* for wide figure, [h] [!] to change the placement
\includegraphics[width=11cm]{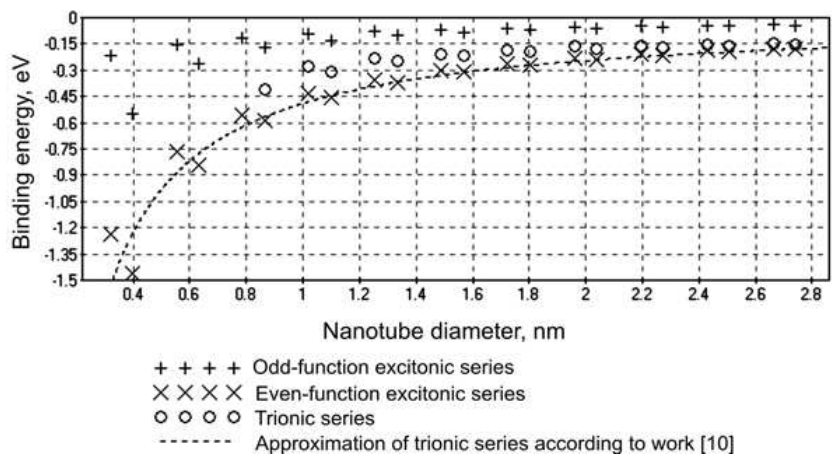}
\vskip-3mm\caption{Dependences of the binding energies for trions
and excitons of both series on the nanotube diameter, in which a
quasiparticle is localized, and the plot of approximation expression
}
\end{figure*}

Notice that the application of the Ritz variational method brings about
a numerical series (in our case, we obtained it in the form of
functions (\ref{1.7}) and (\ref{1.8})), every term of which is
either greater than or equal to the exact solution. Therefore, the
smallest term in the series -- in our case, it is the extremum of
the corresponding function -- is the closest to the exact solution,
approaching it from above.

The choice of functions is rather crude. However, this crudeness in the
choice of trial functions can only elevate the energies of ground states.
Therefore, if the binding energy of an exciton obtained in the framework of
the variational method turns out higher than the energy gap width in the
nanotube, the same relation between those quantities remains valid for the
accurate solution as well.

The function minima lie in the negative region, with the first minimum being
located at a lower energy than the second one. The fact that the energy of
ground states lies in the negative region means that the state is bound,
i.e. it does exist in the case concerned.

In Fig.~1, the dependences of the exciton energy on the diameter of a nanotube, in
which the exciton is localized, are depicted for both series. Note that a
substantial spread in the exciton binding energies for small nanotube
diameters becomes narrower for larger diameters, so that the differences
between excitonic spectra for wide nanotubes are less pronounced than for
narrow ones.

\section{Trionic States}

The trion is a bound state consisting of either two electrons and a hole or
two holes and an electron. In fact, it is an ionized state of the excitonic
molecule (biexciton). The term \textquotedblleft charged
exciton\textquotedblright\ can also be found quite often. For the first
time, the experimental observation of the trionic state in carbon nanotubes was
described in work \cite{matsunaga1}, the results of which we will refer to
below.

In order to find the characteristic values of trion binding energy, we
should solve the corresponding Schr\"{o}dinger equation, analogously to what
was done in the exciton case. To simplify the model, we neglect the
difference between the effective masses of electrons and holes; actually,
their masses differ from each other by less than 3\% \cite{tishchenko1}.
Hence, let the trion mass be simply equal to the effective mass of a hole.
Then, assuming the interaction between each pair of particles composing the
trion to be purely Coulombic, it is possible to write down the Hamiltonian
for a one-dimensional trion in a nanotube in the form
%9
\[
\hat{H}=-\frac{\hbar^2}{2m_{h}}\left[\frac{\partial^2}{\partial
x^2}+\frac{\partial^2}{\partial \xi_1^2}+\frac{\partial^2}{\partial
\xi_2^2}\right]-\frac{e^2}{\sqrt{(x-\xi_1)^2+d^2}}-\]
\begin{equation}\label{1.9}
-\frac{e^2}{\sqrt{(x-\xi_2)^2+d^2}}
+\frac{e^2}{\sqrt{(\xi_1-\xi_2)^2+d^2}},
\end{equation}
where $x$, $\xi _{1}$, and $\xi _{2}$ are the coordinates of
the electron and two holes, respectively; $e$ is the electron charge;
$d$ is a parameter depending on the nanotube diameter (as was done
in the case of excitons, it can be taken equal to the value of
diameter itself); and $m_{h}$ is the trion mass, which coincides in our case
with the hole one.

Actually, any trion includes three two-particle bound states: two excitonic
and one hole--hole. Therefore, it is quite natural to construct the wave
functions of a trion as products of the wave functions that correspond to those
states. This means that the wave function should be tried as a combination of
three Gaussian exponents corresponding to three components describing
two-particle states and a plane wave, the latter describing the motion of
the center of masses of the system as a whole. Specifically,
%10
\[
\Psi_{X^{\pm}}=C_{3}(\xi_1-\xi_2)\exp{[-\alpha \{(x- \xi_1)^{2}+(x-
\xi_2)^{2} \}]} \times \]
\begin{equation}\label{1.10}
\times \exp{[-\beta(\xi_1- \xi_2)^{2}]}\exp{[-i p (x + \xi_1 +
\xi_2)]},
\end{equation}
where $\alpha $ and $\beta $ are small parameters. The presence
of a difference between the hole coordinates, which enters as a
multiplier into the wave function, testifies that the probability
of a state, in which the holes---the components of the trion---come
closer to each other, diminishes and tends to zero, if the holes
approach at an infinitesimally short distance between them.

%Fig. 2
\begin{figure}% figure* for wide figure, [h] [!] to change the placement
\includegraphics[width=6cm]{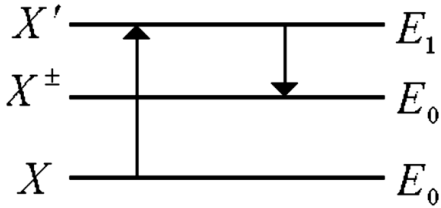}
\vskip-3mm\caption{To generate a trion, an exciton must be
transferred into the excited state, which gives rise to the
appearance of absorption and emission lines similar to those
observed at luminescence. Here, $E_{0}(X)$ is the energy of the
ground excitonic state, $E_{0}(X^{\pm})$ the energy of the ground
trionic state, and $E_{1}(X^{\pm})$ the energy of the first excited
excitonic state  }
\end{figure}

The function to be minimized is not presented here because of its cumbersome
expression, which does not bear any physical information. The minima of the
functions were determined numerically.

\begin{table}[b]
\noindent\caption{Energies of excitonic-family particles for various
zigzag nanotubes and their comparison with the energy gap width:
\boldmath$(n,m)$ are indices of the nanotube chirality, $d$ the nanotube
diameter, $E(X_{\mathrm{even}})$ and $E(X_{\mathrm{odd}})$ the
exciton energies for the even- and odd-function, respectively,
$E(X^{\pm})$ is the trion energy, and $E_{\mathrm{gap}}$ the energy
gap width \cite{tishchenko1}}\vskip3mm\tabcolsep2.2pt
\noindent{\footnotesize\begin{tabular}{c c c c c c}
 \hline \multicolumn{1}{c}
{\rule{0pt}{9pt}$(n, m)$} & \multicolumn{1}{|c}{$d$, nm}&
\multicolumn{1}{|c}{$E(X_{\rm odd})$, eV}&
\multicolumn{1}{|c}{$E(X^{\pm})$, eV}& \multicolumn{1}{|c}{$E(X_{\rm
even})$, eV}&
\multicolumn{1}{|c}{$E_{\rm gap}$, eV}\\%
\hline%
(4,0) &  0.3214 & --1.2434 & --       &  --0.2154 & 2.0749 \\
     (5,0) &  0.3980 & --1.4602 & --       &  --0.5511 & 2.3423 \\
     (7,0) &  0.5526 & --0.7691 & --       &  --0.1530 & 1.3416 \\
     (8,0) &  0.6304 & --0.8392 & --       &  --0.2612 & 1.4153 \\
     (10,0)&  0.7861 & --0.5561 & --       &  --0.1172 & 0.9774 \\
     (11,0)&  0.8641 & --0.5911 & --0.3293 &  --0.1720 & 1.0115 \\
     (13,0)&  1.0202 & --0.4363 & --0.2288 &  --0.0948 & 0.7667 \\
     (14,0)&  1.0983 & --0.4563 & --0.2188 &  --0.1280 & 0.7865 \\
     (16,0)&  1.2546 & --0.3583 & --0.1652 &  --0.0796 & 0.6302 \\
     (17,0)&  1.3328 & --0.3721 & --0.1581 &  --0.1020 & 0.6431 \\
     (19,0)&  1.4892 & --0.3042 & --0.1328 &  --0.0679 & 0.5348 \\
     (20,0)&  1.5674 & --0.3132 & --0.1268 &  --0.0842 & 0.5439 \\
     (22,0)&  1.7238 & --0.2644 & --0.1019 &  --0.0601 & 0.4643 \\
     (23,0)&  1.8020 & --0.2712 & --0.0968 &  --0.0720 & 0.4711 \\
     (25,0)&  1.9585 & --0.2333 & --0.0769 &  --0.0535 & 0.4103 \\
     (26,0)&  2.0367 & --0.2391 & --0.0739 &  --0.0628 & 0.4156 \\
     (28,0)&  2.1932 & --0.2090 & --0.0700 &  --0.0484 & 0.3675 \\
     (29,0)&  2.2715 & --0.2130 & --0.0676 &  --0.0557 & 0.3717 \\
     (31,0)&  2.4280 & --0.1890 & --0.0633 &  --0.0441 & 0.3328 \\
     (32,0)&  2.5062 & --0.1930 & --0.0624 &  --0.0501 & 0.3362 \\
     (34,0)&  2.6628 & --0.1740 & --0.0576 &  --0.0404 & 0.3040 \\
     (35,0)&  2.7410 & --0.1760 & --0.0567 &  --0.0454 & 0.3069 \\
\hline
\end{tabular}}
\end{table}

Zigzag nanotubes, whose diameters are narrower than 1~nm, were not considered because
of the following reason: such narrow nanotubes were not observed as
independent objects, but only as one of the layers in a multiwalled tube.

As was written above, the experiment, in which the lines that could
be identified as trionic were observed for the first time, was made
not earlier than at the beginning of 2011 (see its description in
work \cite{matsunaga1}). In a series of experiments dealing with the
observation of trionic lines, an approximate dependence of the
trionic excitation energy on the nanotube diameter was established.
For a medium with the dielectric permittivity $\varepsilon
_{\mathrm{env}}=3.5$, this dependence looks like $E_{X^{\pm
}}\approx 40/d$. For the sake of comparison with our results, the
following expression bringing about \textquotedblleft
vacuum\textquotedblright\ energies should be used:
%11
\begin{equation}
E_{X^{\pm }}\approx \frac{40}{d}\varepsilon _{\rm env}^{2}.
\label{1.11}
\end{equation}%

\noindent This dependence is plotted in Fig.~1. In agreement with
theoretical predictions, the values estimated with the use of
the variational technique approach from above those following from
the experiment.

An interesting fact consists in that the binding energy of a trion
turned out to have a higher value with respect to the corresponding
exciton energy in the even-function series for all examined
specimens, being at the same time lower than the binding energy of
exciton excitations in the off-function series (Table). Recall that
the excitons in the even-function series correspond to the first
excited state, whereas those in the even-function series to the
ground one. Hence, a trion can emerge as a result of the capture of
a free electron in the conduction band or a hole in the valence band
by excitons, which gives rise to a lowering of the total energy of
the system. The emerging trion is a so-called excimer complex,
because an excited exciton takes part in its generation.

Excimer complexes with the binding energies exceeding the energy of the
ground state cannot exist infinitely long. They should decay within rather a
short time interval by means of an optical or radiationless transition.

We should emphasize that a trion can emerge only if an excited exciton
captures a hole or an electron and lowers its energy. In turn, the excited
exciton can emerge, if an unexcited exciton absorbs energy. Therefore, we
have a three-level luminescence system (Fig.~2). This fact can be used for
the identification of nanotubes and in technological facilities (systems with
pumping).

\section{Conclusions}

Trionic excitation in zigzag SWNTs are shown to be excimers formed, when an
excited exciton captures a free electron or a hole. The excited excitonic
state in SWNTs is unstable with respect to the trionic one. Together with
the instability of a trion with respect to the ground excitonic state, this
fact gives rise to the emergence of a three-level energy system. Taking into
account low concentrations, this circumstance points to a principle
capability of developing an excimer infra-red radiation source, which would
operate in the one-photon mode.

\rezume{%
ОЦІНКА СТАБІЛЬНОСТІ ТРІОННИХ СТАНІВ\\ У ВУГЛЕЦЕВИХ НАНОТРУБКАХ ТИПУ
ЗИГЗАГ}{С. Марченко} {Оцінено стабільності тріонних збуджень у
вуглецевих нанотрубках типу зигзаг. Показано, що тріон є
нестабільним відносно основного екситонного стану та стабільним
відносно збудженого. Отже, тріони в нанотрубках цього типу можуть
бути утворені за рахунок захоплення дірки або електрона екситоном,
що знаходиться у збудженому стані. Тобто, тріон у нанотрубці є
ексимерною сполукою, що приводить до появи трирівневої енергетичної
системи (незбуджений екситон--тріон--збуджений екситон).}

\end{document}